\documentstyle[epsf,12pt]{article}

\begin{document}

\vspace{4cm}

\begin{center}
\Large {\bf POLARIZATION OF VALENCE AND SEA QUARKS IN THE PROTON}
\end{center}

\bigskip
\bigskip
\bigskip
\bigskip

\begin{center}
\large {\bf Wojciech Wi\'slicki}

on behalf of {\bf the Spin Muon Collaboration}

Laboratory for High Energy Physics \\
Soltan Institute for Nuclear Studies \\
Ho\.za 69 \\
PL-00-681 Warsaw \\
email: wislicki@na47sun05.cern.ch 
\end{center}

\vspace{5cm}

{\large {\bf Abstract}}

Analysis was performed of semi-inclusive and inclusive spin asymmetries 
determined from the polarized deep inelastic scattering by the Spin Muon Collaboration.
Combined analysis of data for polarized deuterium and hydrogen targets
allows for separate determination of spin carried by valence {\it u} and {\it d}
quarks and non-strange sea quarks as a function of $x_{Bj}$ in the range 
$0.006<x_{Bj}<0.6$.
It was found that polarization of valence {\it u} quarks is positive and of valence
{\it d} quarks is negative, whereas the sea polarization is small and consistent with
zero within errors.

\vspace{5cm}

Talk presented at XXIXth RENCONTRES DE MORIOND {\it QCD and High Energy Hadronic Interactions},
Meribel, France, 22nd March 1994

To appear in the {\it Proceedings}

\pagebreak

\twocolumn

\Huge {\it T}\normalsize he result of the EMC$^{1)}$, that integrated spin distribution of quarks in the proton 
is far below $\frac{\hbar}{2}$, has provoked intensive further experimentation and 
theoretical activity. Recent experiments at CERN$^{2,3)}$ and SLAC$^{4)}$
determined spin-dependent structure functions $g_1$
of the deuteron, the neutron and the proton thus allowing tests of sum rules for their integrals and 
re-evaluation of the total spin carried by quarks in the nucleon.

One of the hot problems in nucleon structure is the origin of
nucleon spin. In particular, it still remains a mystery 
how spin is shared among valence quarks, sea quarks, gluons and orbital momentum of
nucleon constituents. 
Experimentally these questions can not be answered by using inclusive observables
only. Separation of spin valence and sea components 
is possible when struck quark is tagged by the final state hadron.
For this study, performed within the quark-parton model (QPM), one has to assume the $SU(2)$ isospin
symmetry, charge invariance and factorization of fragmentation functions.
Our analysis aims towards 
determination of polarization of valence and non-strange sea quarks in the nucleon.

We used the polarized muon deep inelastic data from polarized deuterium and hydrogen targets, 
collected by the Spin Muon Collaboration (SMC) at CERN in 1992 and 1993. 
Our experiment 
determines spin-dependent cross section asymmetries. 
The
set up consists of three major elements: the polarized target,
the spectrometer and the polarimeter. 

Polarized $\mu^+$ beam of two energies, 100 GeV and 190 GeV, was used.
Polarization was measured from the shape of the energy spectrum of positrons from
the muon decay and was found to be $-0.820\pm 0.061$ and $-0.803\pm 0.035$ for 100 GeV and
190 GeV, respectively$^{5)}$.

The target consists of two cells filled with
butanol or deuterated butanol and are polarized by dynamic nuclear
polarization in opposite directions. 
Typical values are about 40\% for deuterons and over 80\% for protons. Polarization is determined
with accuracy better than 3\% by measuring the NMR signals with a Q-meter circuits$^{6)}$.
The data are taken simultaneously from both cells and periodically reversing polarization,
so that the beam fluxes and detector acceptances cancel out in measured asymmetries.
\begin{figure}[t]
 \begin{center}
  \vskip -1.cm
  \mbox{\epsfxsize=7cm\epsfysize=7cm\epsffile{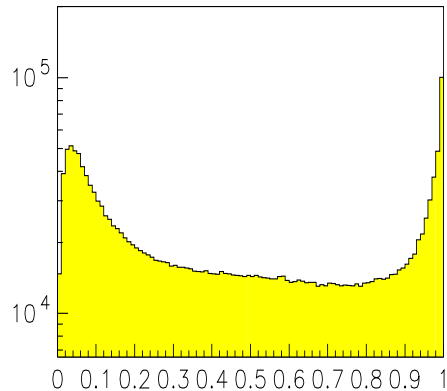}}
 \end{center}
 \vskip -1.2cm
 \caption{\em $E_{em}/(E_{em}+E_{had})$}
\end{figure}

Particles are detected and measured in the forward spectrometer$^{7)}$.
Scattered muon is defined as
a particle penetrating the iron wall and giving a signal in streamer tubes
behind it. 
All charged particles are tracked in about 150 planes of proportional and streamer chambers
and used to determine the interaction vertex.
Their momenta are measured by using the bending magnet. 
High energy electrons from conversion of radiative photons
can occasionally fit to the $\vec{\mu}\vec{N}$ interaction vertex.
A calorimeter was used to discard such electrons thus avoiding their misidentification
as final state hadrons. 
\begin{figure}[h]
 \begin{center}
 \vskip -15mm
  \mbox{\epsfxsize=7cm\epsfysize=7cm\epsffile{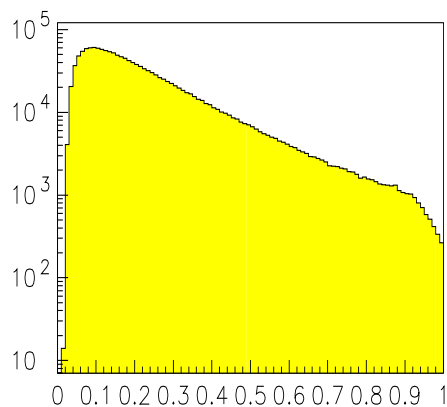}}
 \end{center}
 \vskip -1.2cm
 \caption{\em $z_{had}=E_h/\nu$}
 \vskip -4mm
\end{figure}
The hadron is operationally defined as a particle 
fulfilling acceptance requirements of the calorimeter, for which
the ratio of energy loss in the electromagnetic part ($E_{em}$)  to the total
energy deposited in electromagnetic and hadronic ($E_{had}$) parts does not exceed 80\%.
The spectrum of this ratio is displayed in fig. 1. 
Although the kinematically accessible region of $x_{Bj}$ is somewhat wider for 190 GeV data than for 100 GeV,
the combined analysis is possible in the overlap range of $0.006<x_{Bj}<0.6$. For both samples 
$Q^2>1$ GeV$^2$ was required.
\begin{figure}[t]
 \begin{center}
 \vskip -1.cm
  \mbox{\epsfxsize=7cm\epsfysize=5cm\epsffile{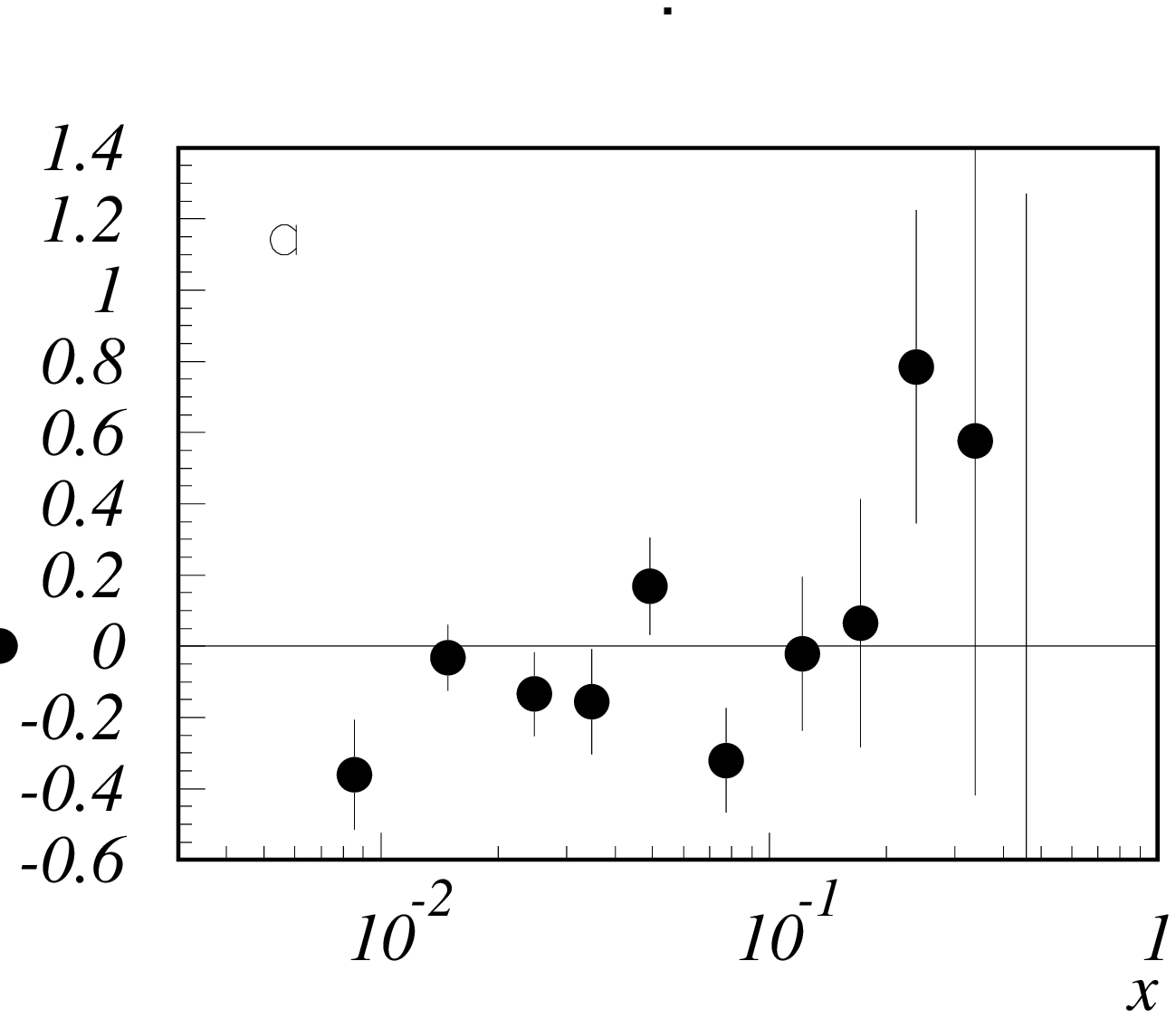}}
  \vskip -10mm
  \mbox{\epsfxsize=7cm\epsfysize=5cm\epsffile{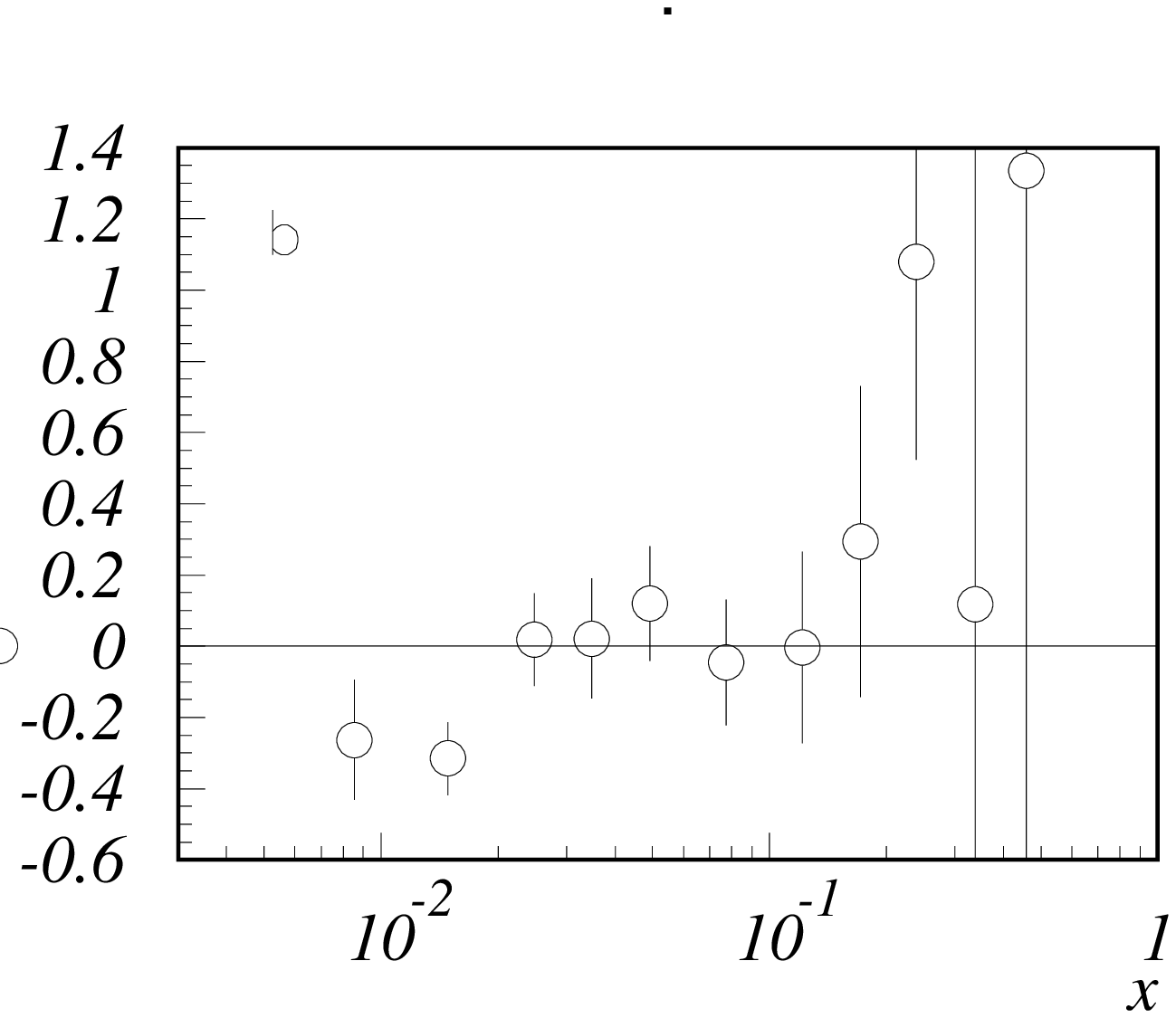}}
 \end{center}
 \vskip -1.cm
 \caption{\em Semi-inclusive asymmetries of spin-dependent cross sections for muoproduction of positive (a)
  and negative (b) hadrons on deuterons at 100 GeV}
 \vskip -4mm
\end{figure}
Due to limited angular acceptance of our spectrometer we poorly accept
hadrons with $z_{had}=E_h/\nu$ below 0.1 (cf. fig. 2). In this analysis a cut on $z>0.2$ was applied.
This choice is a compromise
between the most effective tagging of the struck quark for semi-inclusive asymmetries and non-dramatic
statistics loss.
After applying kinematic and geometric cuts we are left with comparable samples of $4.5\times 10^6$
deep inelastic events for each, 100 GeV deuterium and 190 GeV hydrogen 
data set, where $\langle Q^2\rangle$ was equal to 4.6 and 10 GeV$^2$, respectively. 
Corresponding hadron samples amount to $1.6\times 10^6$ and $1.4\times 10^6$ charged hadrons.
Spin asymmetries are measured for virtual photon deep inelastic scattering cross section on 
polarized proton (p) and deutron (d), $A_{d(p)}^{\mu}$, and for muoproduction of charged positive
(+) or negative (-) hadrons, $A_{d(p)}^{+(-)}$, 
\begin{equation}
A_{d(p)}^{\mu(+(-))}=\alpha\frac{\sigma_{d(p)_{1/2}}^{\mu(+(-))}-\sigma_{d(p)_{3/2}}^{\mu(+(-))}}
                                {\sigma_{d(p)_{1/2}}^{\mu(+(-))}+\sigma_{d(p)_{3/2}}^{\mu(+(-))}},
\end{equation}
where indices $1/2$ and $3/2$ refer to the total spin projection in the direction of virtual photon.
For deuteron the cross sections in (1) are assumed to be the average of proton and neutron cross sections. 
The factor $\alpha=1-\frac{3}{2}\omega_D$ accounts for $\omega_D\simeq0.06$  
probability of deuteron to be in D-state and
for the proton $\alpha=1$. The cross sections depend on polarized and unpolarized quark distributions
and, for semi-inclusive asymmetries, on quark fragmentation functions.
\begin{figure}[h]
 \begin{center}
 \vskip -1.cm
  \mbox{\epsfxsize=7cm\epsfysize=5cm\epsffile{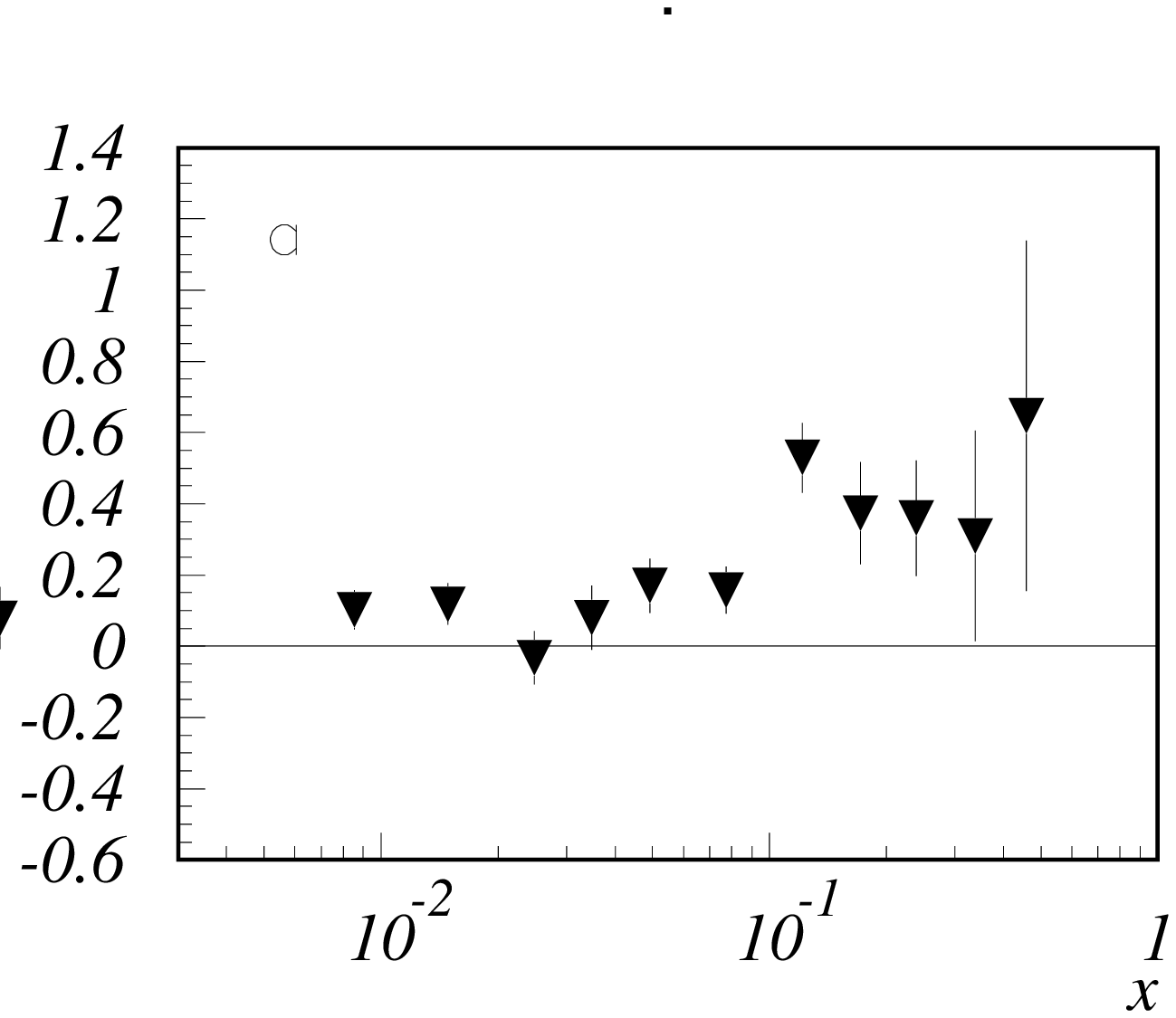}}
  \vskip -10mm
  \mbox{\epsfxsize=7cm\epsfysize=5cm\epsffile{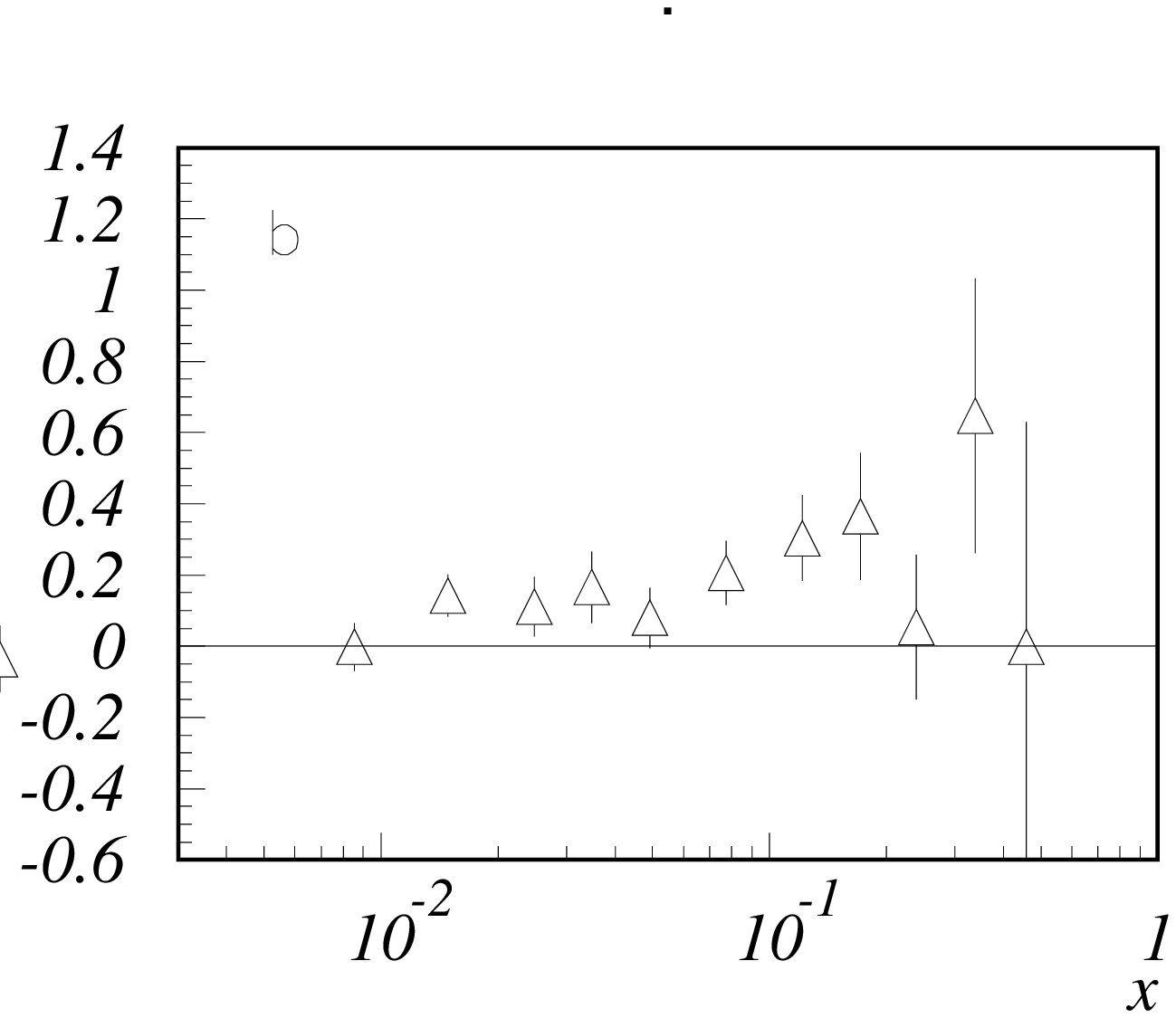}}
 \end{center}
 \vskip -1.cm
 \caption{\em Semi-inclusive asymmetries of spin-dependent cross sections for muoproduction of positive (a)
  and negative (b) hadrons on protons at 190 GeV}
\end{figure}

Experimentally the asymmetries (1) are determined from the numbers of
events (inclusive) or charged
particle yields (semi-inclusive) taken for two beam-target spin configurations and accounting for
the degree of polarization of beam and target, the amount of unpolarizable material in the target (dilution
factor), virtual gamma depolarization and radiative corrections.

Inclusive asymmetries for deuteron
and proton were published$^{2,3)}$ and semi-inclusive asymmetries are displayed in fig. 3 for positive
and negative hadrons from the deuteron and in fig. 4 from the proton. 

The errors for experimental points presented
in these and the following figures are statistical only.
\begin{figure}[t]
 \begin{center}
 \vskip -1.cm
  \mbox{\epsfxsize=7cm\epsfysize=5cm\epsffile{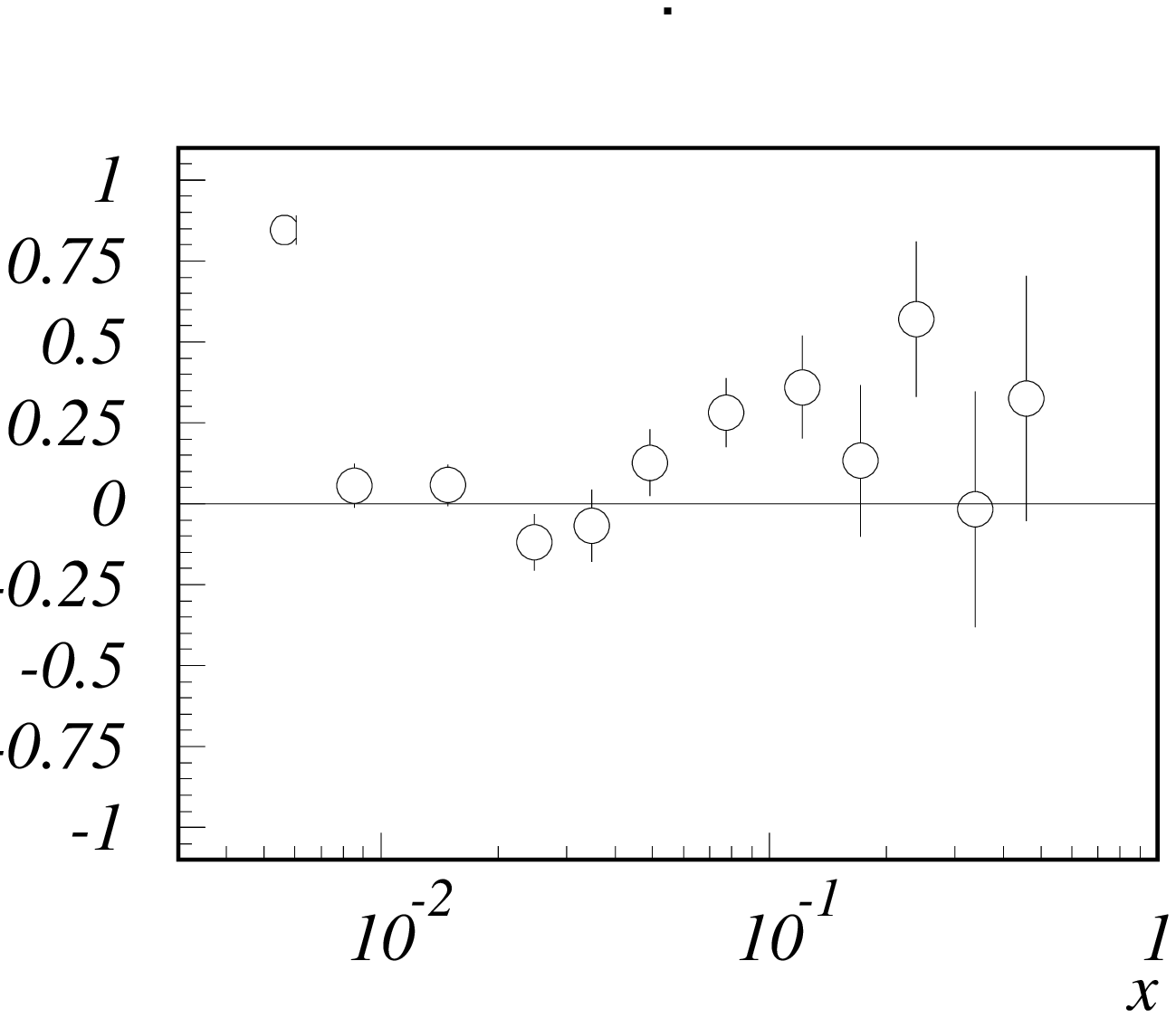}}
  \vskip -10mm
  \mbox{\epsfxsize=7cm\epsfysize=5cm\epsffile{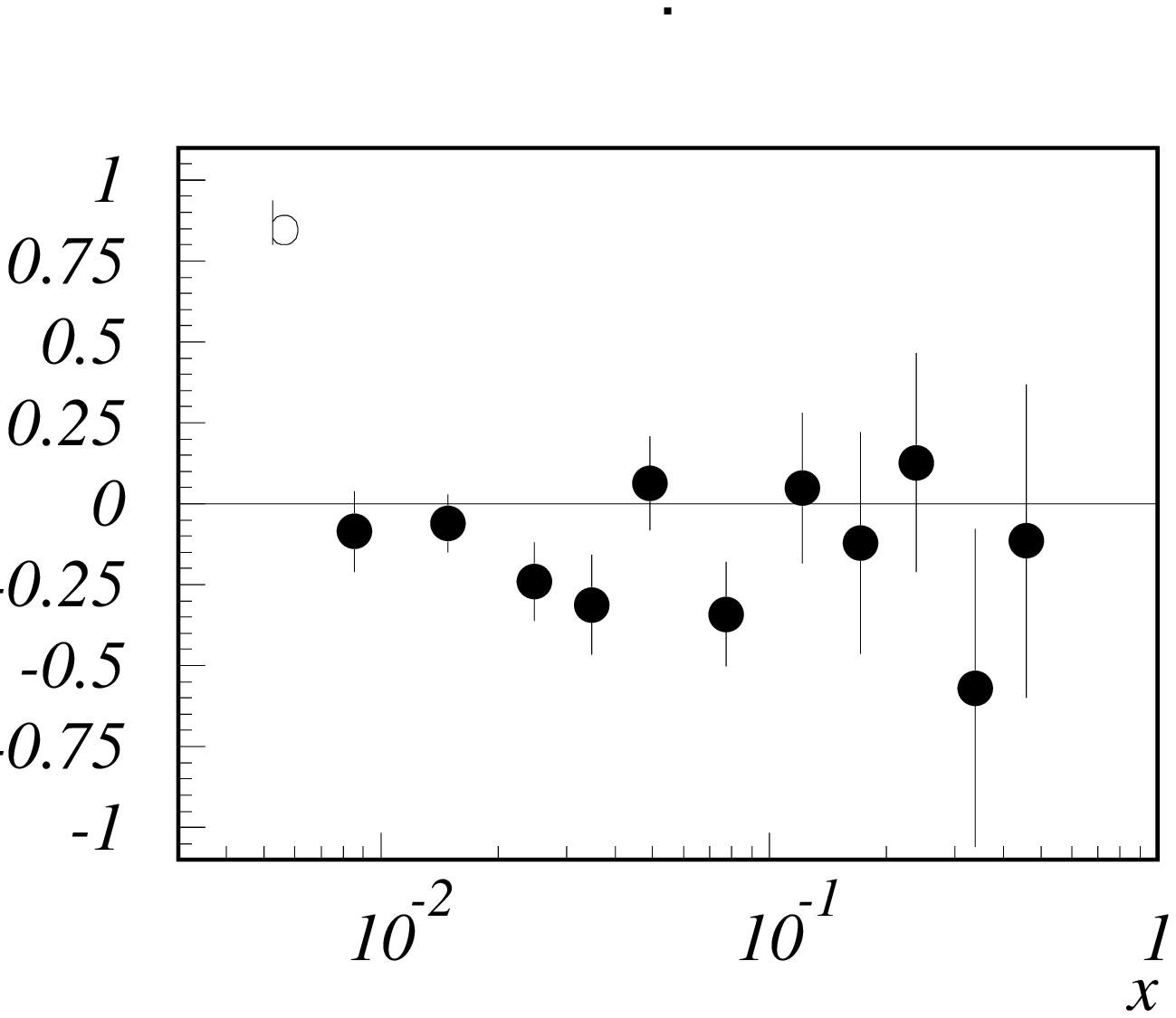}}
  \vskip -10mm
  \mbox{\epsfxsize=7cm\epsfysize=5cm\epsffile{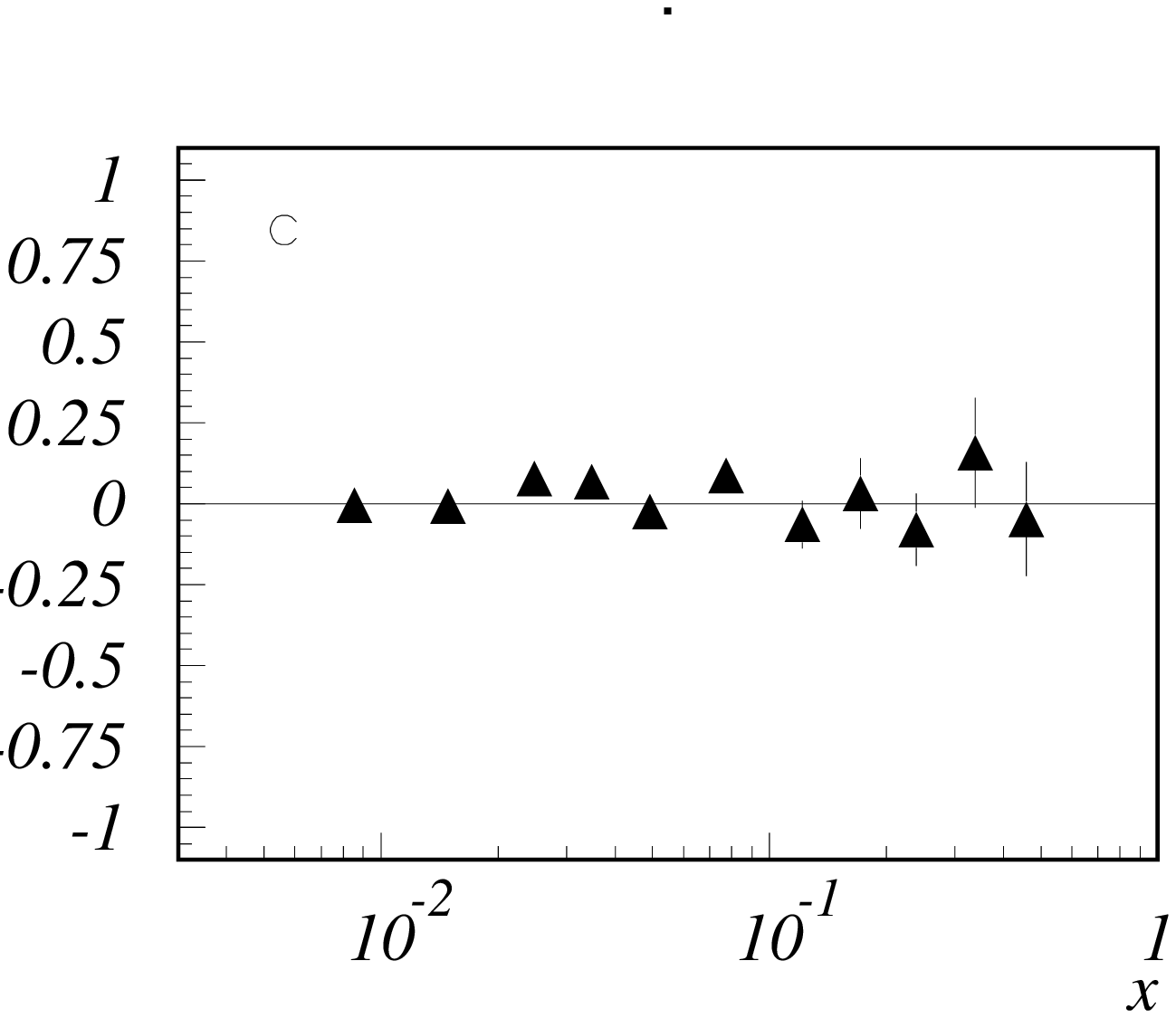}}
 \end{center}
 \vskip -1.cm
 \caption{\em Quark spin distribution functions $x\Delta u_v(x)$ (a), $x\Delta d_v(x)$ (b) and $x\Delta\bar{q}(x)$ (c)
              at $Q^2=10$ GeV$^2$}
 \vskip -5mm
\end{figure}

In the framework of the QPM all asymmetries can be expressed as linear
combinations of polarizations of valence, $\Delta u_v$ and $\Delta d_v$, non-strange sea, $\Delta \bar{q}$, 
and strange, $\Delta s$, quarks,
\begin{eqnarray}
A_{d,p}^{\mu+-} = & \nonumber \\
c_{1_{d,p}}^{\mu+-}\Delta u_v+c_{2_{d,p}}^{\mu+-}\Delta d_v+c_{3_{d,p}}^{\mu+-}\Delta \bar{q}
+c_{4_{d,p}}^{\mu+-}\Delta s,
\end{eqnarray}
where the $SU(2)$ isosymmetry was assumed for the non-strange sea, i.e. $\Delta \bar{u}=\Delta \bar{d}\equiv\Delta \bar{q}$, 
$\Delta$ being the difference between distributions of spins parallel and antiparallel to the nucleon spin. 
Following the usual convention all quark distributions refer to the proton; i.e.
in formulae
for deuteron asymmetries it was assumed that distribution of $u_v(d_v)$ in neutron is the same as
$d_v(u_v)$ for proton. 
\begin{figure}[t]
 \begin{center}
 \vskip -1.cm
  \mbox{\epsfxsize=7cm\epsfysize=5cm\epsffile{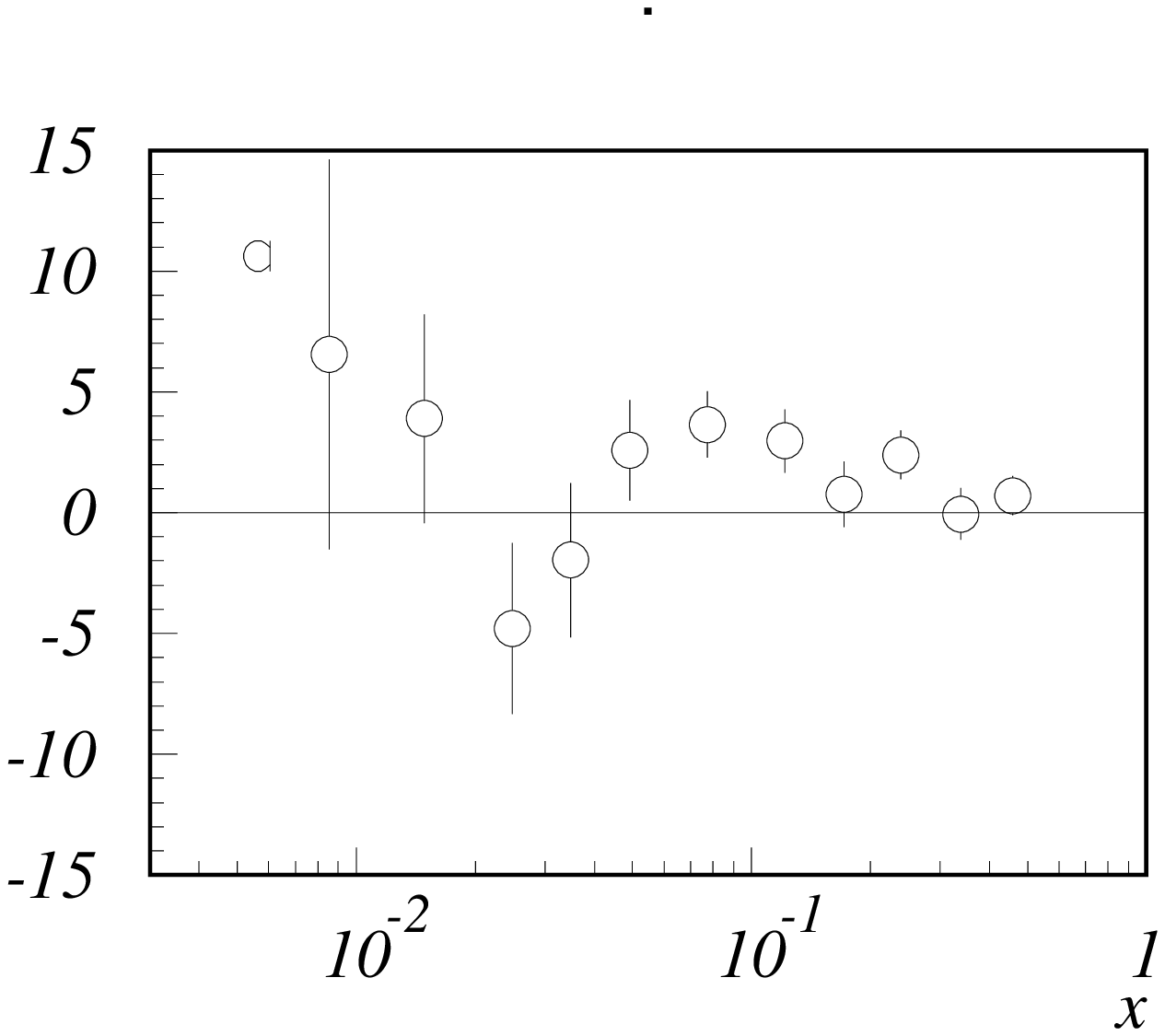}}
  \vskip -10mm
  \mbox{\epsfxsize=7cm\epsfysize=5cm\epsffile{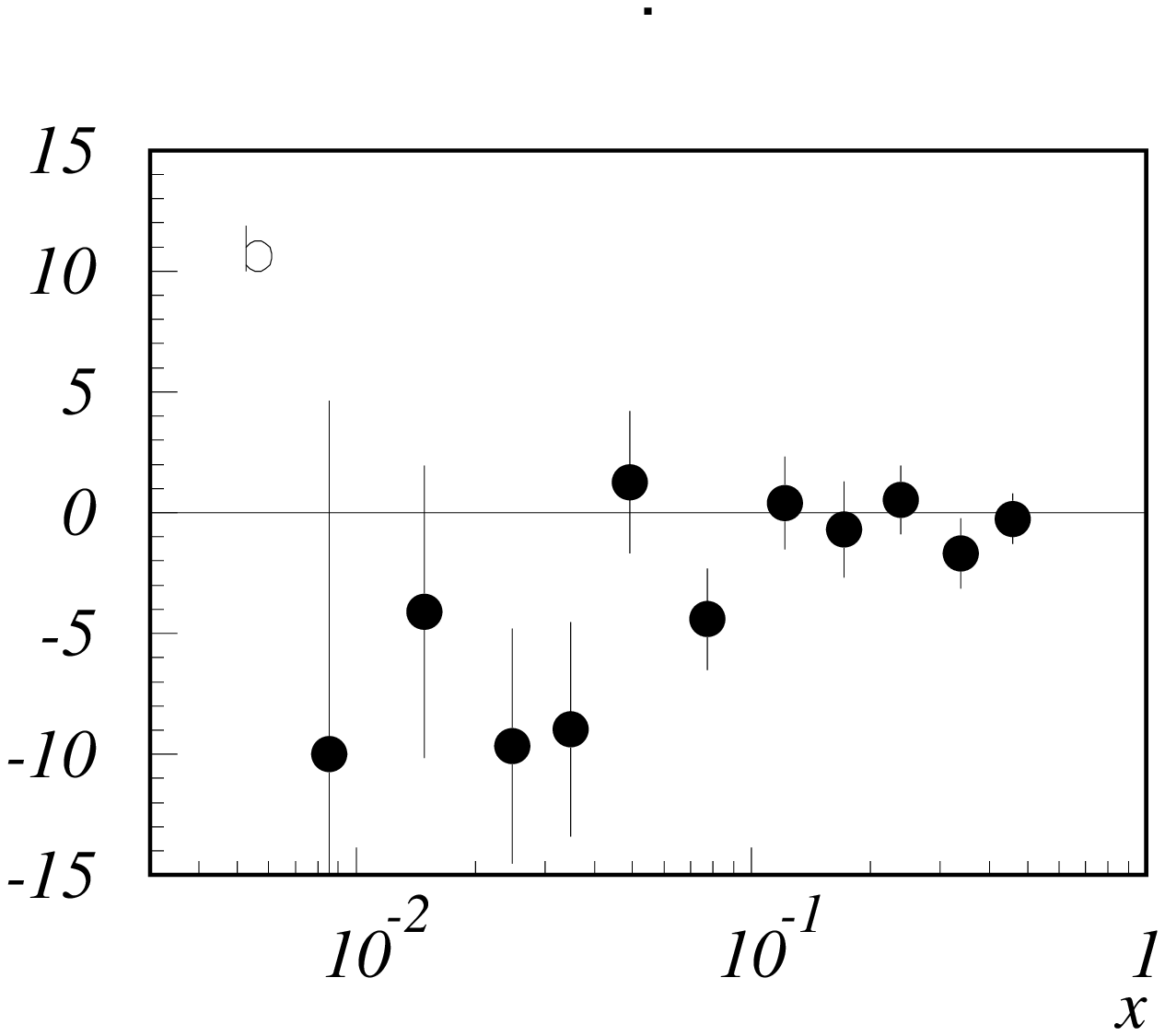}}
  \vskip -10mm
  \mbox{\epsfxsize=7cm\epsfysize=5cm\epsffile{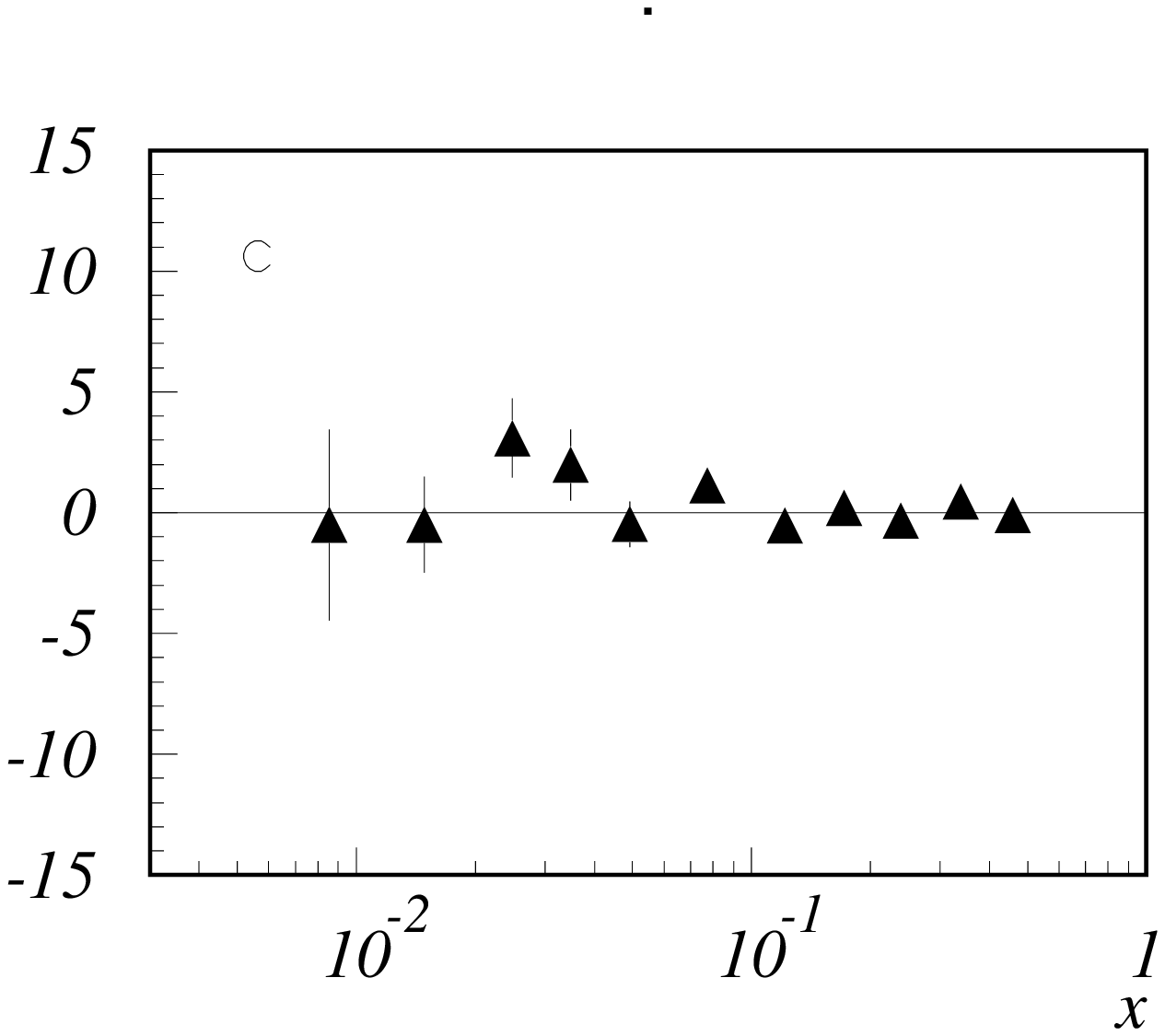}}
 \end{center}
 \vskip -10mm
 \caption{\em Quark spin distribution functions $\Delta u_v(x)$ (a), $\Delta d_v(x)$ (b) and $\Delta\bar{q}(x)$ (c)
              at $Q^2=10$ GeV$^2$}
 \vskip -5mm
\end{figure}
Coefficients $c_{i_{d,p}}^{\mu+-}$ depend on unpolarized quark distribution functions, for which the 
parametrizations were used$^{8)}$. For semi-inclusive asymmetries, in addition, they depend also
on quark fragmentation functions
into hadrons, integrated over $z_{had}$ from $0.2$ to 1, for which we used the EMC measurements$^{9)}$.
It was assumed that the fragmentation functions are invariants of charge and isospin transformations.
For example, for pions, the most abundant
hadrons in the final state, the favoured fragmentation functions are 
\begin{equation}
 D_u^{\pi^+} = D_{\bar{d}}^{\pi^+} = D_d^{\pi^-} = D_{\bar{u}}^{\pi^-} 
\end{equation}
and the unfavoured fragmentation functions are
\begin{equation}
 D_d^{\pi^+} = D_{\bar{u}}^{\pi^+} = D_u^{\pi^-} = D_{\bar{d}}^{\pi^-}. 
\end{equation}
\begin{table}[h]
\vskip -12mm
\caption{\em Integrals of quark polarizations over measured range $0.006<x<0.6$}
\vskip -5mm
\begin{center}
\begin{tabular}{|c|c|} \hline 
 & $\int_{0.006}^{0.6}\Delta q(x)dx$\\ \cline{1-2}
$\Delta u_v$ & $0.767\pm 0.250$\\ \cline{1-2}
$\Delta d_v$ & $-0.615\pm 0.330$\\ \cline{1-2}
$\Delta \bar{q}$ & $0.064\pm 0.060(0.126)$\\ \hline 
\end{tabular}
\end{center}
\end{table}
\begin{table}[h]
\vskip -12mm
\caption{\em Integrals of quark polarizations over unmeasured regions of $x<0.006$ and $x>0.6$}
\vskip -5mm
\begin{center}
\begin{tabular}{|c|c|c|} \hline 
 & $\int_{0}^{0.006}\Delta q(x)dx$ & $\int_{0.6}^{1}\Delta q(x)dx$\\ \cline{1-3}
$\Delta u_v$ & $0.001\pm 0.009$ & $0.061\pm 0.072$\\ \cline{1-3}
$\Delta d_v$ & $-0.032\pm 0.013$ & $-0.048\pm 0.047$\\ \cline{1-3}
$\Delta \bar{q}$ & $0.004\pm 0.004$ & $0.001\pm 0.106$\\ \hline
\end{tabular}
\end{center}
\end{table}
\begin{table}[h]
\vskip -12mm
\caption{\em Integrals of quark polarizations over the range of $0<x<1$}
\vskip -5mm
\begin{center}
\begin{tabular}{|c|c|} \hline 
 & $\int_0^1 \Delta q(x)dx$\\ \cline{1-2}
$\Delta u_v$ & $0.829\pm 0.260$\\ \cline{1-2}
$\Delta d_v$ & $-0.673\pm 0.333$\\ \cline{1-2}
$\Delta \bar{q}$ & $0.068\pm 0.068(0.126)$\\ \hline
\end{tabular}
\end{center}
\vskip -5mm
\end{table}

Polarization of strange quarks can not be determined from eqns. (2), because its contribution to asymmetries is
much smaller than the non-strange quarks. 
We assumed that the strange quark spin is distributed in $x$ in the same way as unpolarized $s(x)$
obtained from the CCFR$^{12)}$ and that the integrated $\Delta s$ is equal to $-0.12$, as
determined in ref. 3. 
The overdetermined system of 6 equations (2), consisting of semi-inclusive asymmetries for positive and negative 
hadrons and one inclusive asymmetry for each target, was solved by the minimum $\chi^2$ method.
Resulting quark polarization distributions $x\Delta q$ are presented in fig. 5. 
The same result, but without momentum weight, $\Delta q$, is shown in fig. 6.
Values in figs. 5, 6 are evolved to common $Q^2=10$ GeV$^2$. In this procedure it was
assumed that asymmetries do not depend on $Q^2$, as supported by our observation$^{10)}$. Unpolarized
quark distribution functions and fragmentation functions were evolved from the $\langle Q^2\rangle$ measured in given $x$
bin to 10 GeV$^2$. Inspection of figs. 5 and 6 reveals that valence {\it u} quarks are polarized in the
direction of proton spin and valence {\it d} quarks are polarized in opposite way. Positive polarization of {\it u}
quarks and negative of {\it d} quarks is expected
from values of the $SU(3)$ coupling constants $F$ and $D$, which constrain the integrals of 
$\Delta u$ and $\Delta d$$^{1)}$,
but the $x$-dependence of valence components is measured for the first time. It is seen
from figs. 5c and 6c that the non-strange sea in the proton does not exhibit significant polarization over the
measured range of $x_{Bj}$. 

Behaviour of polarized structure functions below our measured $x$
is a matter of debate$^{11)}$. 
Since for $x<0.04$ 
we do not observe any significant deviation of $\Delta q$ from a constant, neither for 
the valence nor for the sea,
we obtained its value from a fit in the range $0.006<x<0.04$ and  extrapolated it below 
$x=0.006$, down to $x=0$.
This type of $x$-dependence is consistent with Regge behaviour $x^{\alpha}$ with
$\alpha=0$, as we assumed in our analysis of the $g_1$ structure functions$^{2,3)}$.
Tiny amount of spin from the low-$x$ extrapolation was added to our integral, as listed
in table 2. 

To estimate the contribution from $x>0.6$ a fit was performed of the function $Ax^B(1-x)^C$ 
for $0.1<x<0.6$ and extrapolation beyond $x=0.6$ gave the values displayed in table 2.

The main contribution to the error of $\Delta \bar{q}$ comes from the region of high $x$, where statistical
precision of our data is worse than for small $x$, as seen from fig. 5c. 
The error for $\Delta \bar{q}$,
as calculated with errors on points above $x=0.1$, is given in parentheses in lower
rows of tables 1 and 3. 
On the other hand, we observe that all points for $x>0.1$ are consistent with zero within one standard
deviation and their mean value is zero. The integral of unpolarized sea in this region amounts to
$0.035$, roughly a quarter of our statistical error. Thus, we can make our conclusion about 
$\Delta \bar{q}$ more stringent, by assuming the maximum error for $x>0.1$ being equal to $0.035$, 
which gives the overall error on $\Delta \bar{q}$ equal to $0.068$, as given in tables 1 and 3.

In fig. 7 the integrals $\int_{x_{min}}^1\Delta q$ are displayed, and the
values of integrals over the full $x$ range are denoted by asterisks . Corresponding 
numbers for the measured region of $x$, unmeasured region and the full range are listed in tables 1, 2 and 3. 

In solving eqns. (2) we found that sensitivity of non-strange quark polarizations on $\Delta s(x)$ is negligible,
because $c_4\ll c_{1,2,3}$.
\begin{figure}[t]
 \begin{center}
 \vskip -15mm
  \mbox{\epsfxsize=7cm\epsfysize=5cm\epsffile{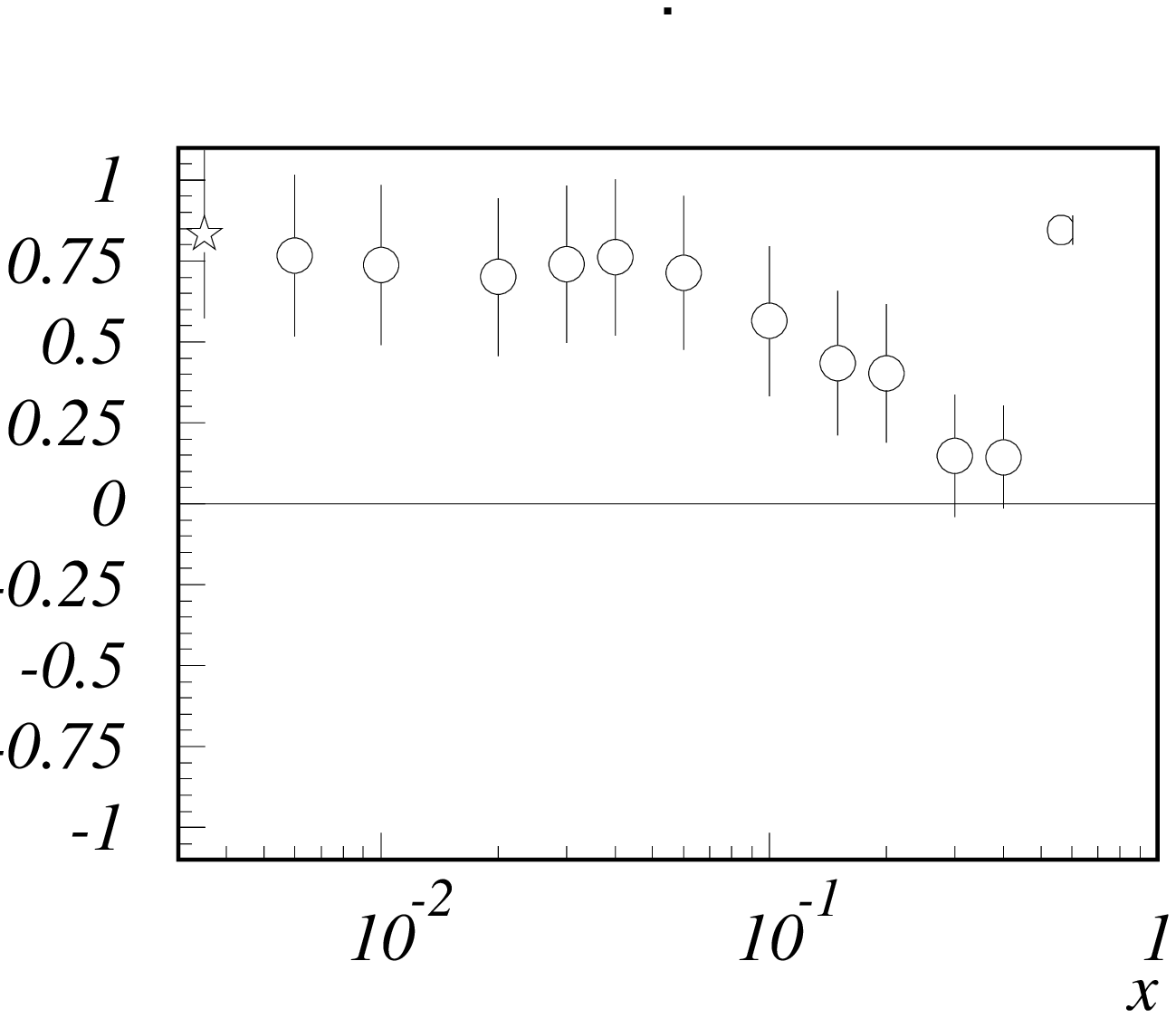}}
  \vskip -10mm
  \mbox{\epsfxsize=7cm\epsfysize=5cm\epsffile{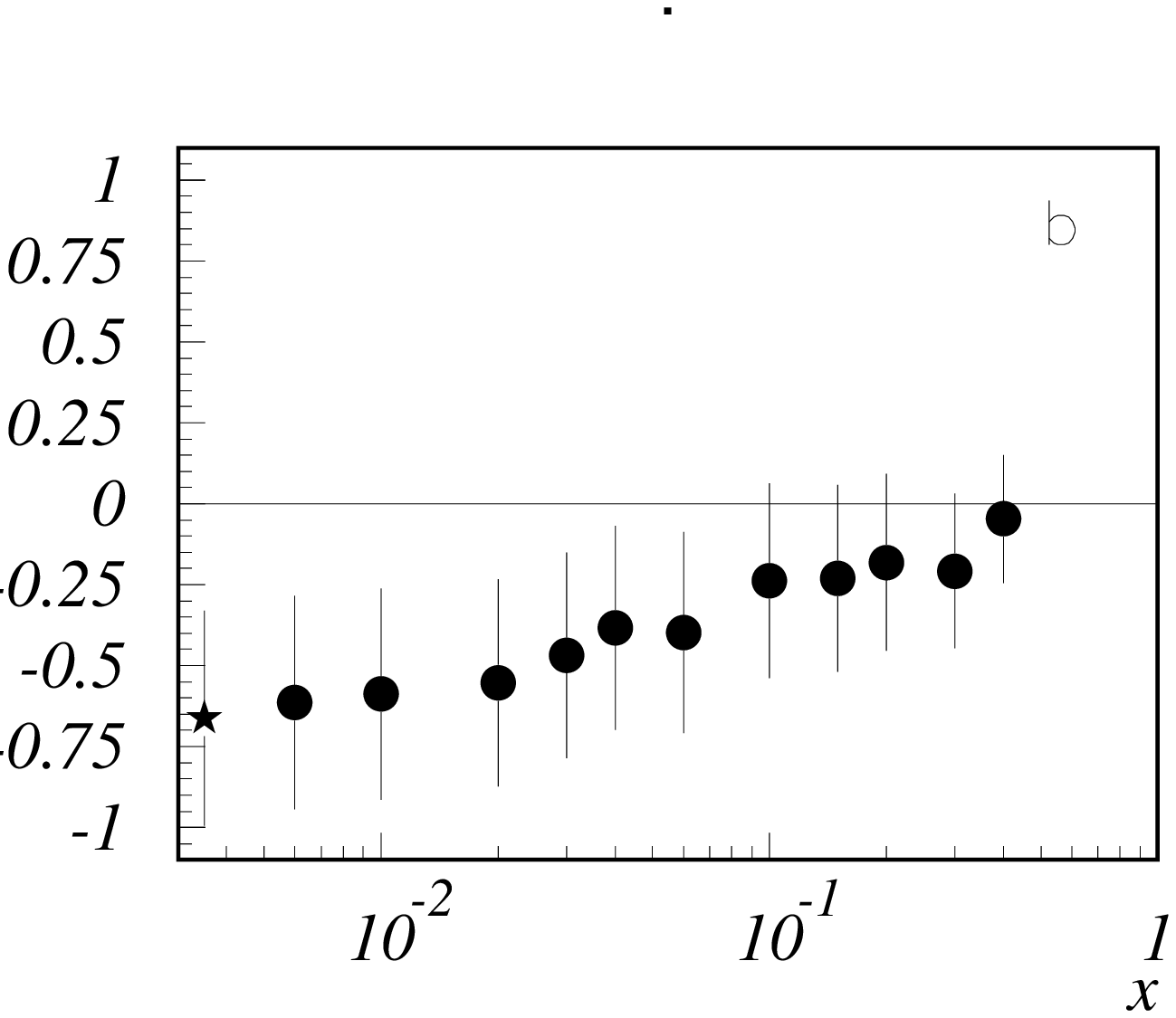}}
  \vskip -10mm
  \mbox{\epsfxsize=7cm\epsfysize=5cm\epsffile{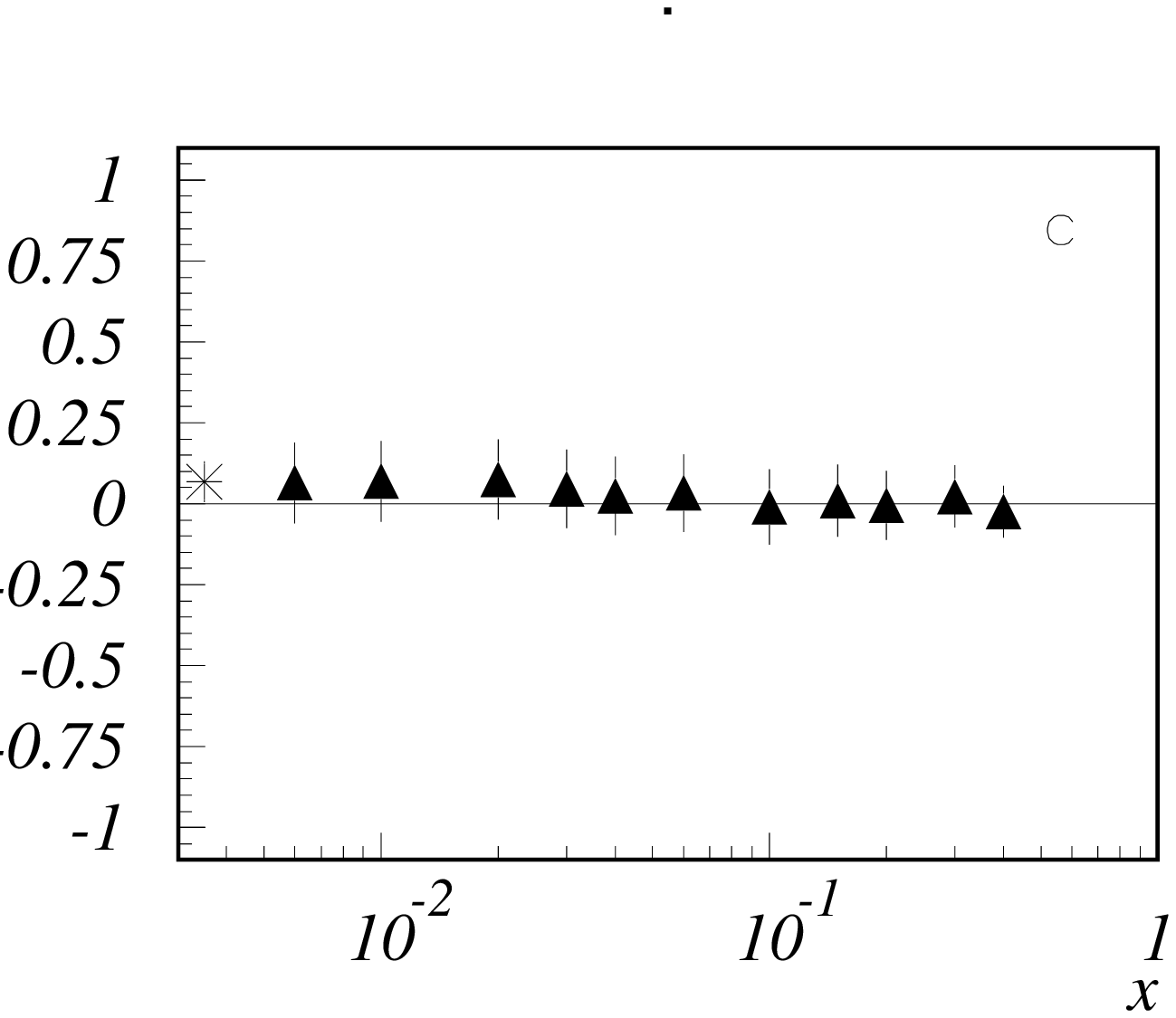}}
 \end{center}
 \vskip -1.cm
 \caption{\em The values of integrals 
$\int_{x_{min}}^1 \Delta q(x)dx$ for $\Delta u_v$ (a), $\Delta d_v$ (b) and $\Delta \bar{q}$ (c).
Asterisk for each figure denotes the value of the integral for 
$x_{Bj}$ from 0 to 1 and includes contribution from extrapolation to unmeasured regions. Errors are
explained in the text.}
\vskip -5mm
\end{figure}
We varied
$\Delta s$ between 0 and $-0.12$
and we found that integrals of quark polarizations change by no more than 1.6\%.

Concluding, we measured the spin distribution of the valence quarks and the non-strange sea quarks
in the proton. We found that $u_v$ quarks are polarized parallel and $d_v$ quarks - antiparallel
to the proton spin, whereas the non-strange sea carries small amount of spin, consistent with zero
within our errors. It is highly desirable to continue the study of sharing of the nucleon spin
between different degrees of freedom in present$^{2,3,4)}$ and \linebreak
proposed$^{13,14,15)}$ experiments.

\vspace{5mm}

{\large {\bf Bibliography}}\\
$[1]$ EMC, J.Ashman et al., Phys.Lett. {\bf B206}\linebreak
\hspace*{7mm}(1988)364 and Nucl.Phys. {\bf B328}(1989)1\\
$[2]$ SMC, B.Adeva et al., Phys.Lett. {\bf B302}\linebreak
\hspace*{7mm}(1993)533\\
$[3]$ SMC, D.Adams et al., CERN-PPE/94-57,\linebreak 
\hspace*{7mm}submitted to Phys.Lett.{\bf B} and, also,\linebreak
\hspace*{7mm}presented by Peter Shanahan at this\linebreak 
\hspace*{7mm}conference\\
$[4]$ SLAC E142, D.L.Anthony et al., Phys.Rev.\linebreak
\hspace*{7mm}Lett. {\bf 71}(1993)959\\
$[5]$ SMC, B.Adeva et al., Nucl.Instr.Meth.\linebreak 
\hspace*{7mm}{\bf A343}(1994)363\\
$[6]$ SMC, B.Adeva et al., CERN-PPE/94-54,\linebreak 
\hspace*{7mm}to appear in Nucl.Instr.Meth. {\bf A}\\
$[7]$ EMC, J.P.Albanese et al., Nucl.Instr.Meth.\linebreak
\hspace*{7mm}{\bf 212}(1983)111\\
$[8]$ A.Donnachie and P.V.Landshoff, Z.Phys.\linebreak
\hspace*{7mm}{\bf C61}(1994)139 and M.Gluck et al.\linebreak
\hspace*{7mm}Phys.Lett. {\bf B306}(1993)391\\
$[9]$ EMC, M.Arneodo et al., Nucl.Phys. {\bf B321}\linebreak
\hspace*{7mm}(1989)541\\
$[10]$ SMC, B.Adeva et al., Phys.Lett. {\bf B320}\linebreak
\hspace*{7mm}(1994)400\\
$[11]$ J.Kuti, MIT rep. CTP-234, 1971,\linebreak
\hspace*{7mm}R.Heimann, Nucl.Phys. {\bf B64}(1973)429,\linebreak
\hspace*{7mm}F.E.Close and R.G.Roberts, Phys.Rev.Lett.\linebreak 
\hspace*{7mm}{\bf 60}(1988)1471,\\
\hspace*{7mm}J.Ellis and M.Karliner, Phys.Lett. {\bf B213}\linebreak
\hspace*{7mm}(1988)73,\\
\hspace*{7mm}G.Preparata et al., Phys.Lett. {\bf B231}\linebreak
\hspace*{7mm}(1989)483\\
$[12]$ CCFR, S.A.Rabinowitz et al., Phys.Rev.\linebreak
\hspace*{7mm}Lett. {\bf 70}(1993)134\\
$[13]$ E-143, E-154 and E-155 SLAC\linebreak 
\hspace*{7mm}Proposals (1989-1993)\\
$[14]$ HERMES Coll., K.Coulter et al.,\linebreak 
\hspace*{7mm}DESY/PRC 90/1(1990)\\
$[15]$ R.A.Kunne et al., LNS/Th/93-1(1993)

\end{document}